\documentclass{article}

\usepackage{authblk}
\usepackage[colorlinks,urlcolor=blue,linkcolor=blue,citecolor=blue]{hyperref}
\usepackage{booktabs}
\usepackage{graphicx}
\usepackage{comment} % Comment blocks
\usepackage{subcaption}
\usepackage{marginnote}
\usepackage{tablefootnote}
\usepackage{multirow}
\usepackage{balance}
\begin{document}

\title{The Making of a Community Dark Matter Dataset with the National Science Data Fabric}

\author[1]{Amy Roberts}
\author[3]{Jack Marquez}
\author[3]{Kin Hong NG}
\author[1]{Kitty Mickelson}
\author[2]{Aashish Panta}
\author[2]{Giorgio Scorzelli}
\author[2]{Amy Gooch}
\author[4]{Prisca~Cushman}
\author[4]{Matthew~Fritts}
\author[4]{Himangshu~Neog}
\author[2]{Valerio~Pascucci}
\author[3]{Michela Taufer}

\affil[1]{University of Colorado Denver}
\affil[2]{University of Utah}
\affil[3]{University of Tennessee Knoxville}
\affil[4]{School of Physics and Astronomy, University of Minnesota, Minneapolis, MN 55455}

\maketitle

\begin{abstract}
Dark matter is believed to constitute approximately 85\% of the universe’s matter, yet its fundamental nature remains elusive. 
Direct detection experiments, though globally deployed, generate data that is often locked within custom formats and non-reproducible software stacks, limiting interdisciplinary analysis and innovation. 
This paper presents a collaboration between the National Science Data Fabric (NSDF) and dark matter researchers to improve accessibility, usability, and scientific value of a calibration dataset collected with Cryogenic Dark Matter Search (CDMS) detectors at the University of Minnesota. 
We describe how NSDF services were used to convert data from a proprietary format into an open, multi-resolution IDX structure; develop a web-based dashboard for easily viewing signals; and release a Python-compatible CLI to support scalable workflows and machine learning applications. 
These contributions enable broader use of high-value dark matter datasets, lower the barrier to entry for new collaborators, and support reproducible, cross-disciplinary research.
\end{abstract}

\section{Introduction to Dark Matter Searches}

Visible matter at the galactic scale~\cite{VeraRubin, Villano2023, RevModPhys.90.045002} and in groups of galaxies~\cite{Zwicky2009-jw, Neyman1961-jn, RevModPhys.90.045002} moves too fast to be held together by the gravitational pull of visible matter alone. 
This discrepancy suggests that there is additional, unseen matter, commonly referred to as ``dark matter," providing the extra gravitational attraction needed to prevent these structures from flying apart. 
At the cosmic scale, a slow-moving mass that interacts only gravitationally helps us accurately describe the Cosmic Microwave Background~\cite{Komatsu2011-py, Planck}. 
Dark matter is estimated to constitute approximately 85\% of the total matter in the universe~\cite{Planck}, yet its fundamental properties remain unknown. Determining its nature could resolve long-standing discrepancies observed at galactic, cluster, and cosmological scales~\cite{osti_1320565}.

One leading class of theories postulates that dark matter is composed of yet-undiscovered particles that interact weakly with ordinary matter~\cite{RevModPhys.90.045002}. To detect these elusive interactions, direct detection experiments aim to observe rare scattering events between dark matter particles and atomic nuclei in ultra-sensitive detectors. 
More than 50 such experiments are currently operational worldwide. However, no confirmed dark matter detection has yet been made, making this one of the most critical and active open problems in physics.
These experiments must operate in environments with extremely low background radiation to detect such rare events. This is achieved by: (1) using low-radioactivity materials in the detector’s construction, (2) placing detectors underground to shield them from cosmic radiation, and (3) designing sensors that can differentiate between interactions involving electrons and those involving atomic nuclei.

Detectors are generally optimized to probe a specific dark matter mass range, resulting in a diversity of detector technologies across experiments. 
Nevertheless, many experiments share common data characteristics: they record time-series signals from multiple sensors and use distinct sensor types to distinguish background electron events from potential dark matter-induced nuclear events. 
These shared technical requirements create an opportunity for common tools and methods across the field. 
Yet, the lack of standardized software stacks and consistent data formats makes cross-collaboration and reproducibility challenging.

This paper addresses these barriers by presenting technical contributions that increase access and usability of dark matter data. Specifically, we introduce an open-access calibration dataset collected at the University of Minnesota using Cryogenic Dark Matter Search (CDMS)-style sensors. This dataset was released through a collaboration involving researchers from the University of Colorado Denver, the University of Minnesota, the University of Tennessee, Knoxville, and the University of Utah as part of the National Science Data Fabric (NSDF) initiative.

Through this partnership, we addressed three key challenges: 
(1) improving data accessibility by transforming custom experimental formats into interoperable structures using open-source tools; 
(2) shortening the learning curve with an interactive, web-based dashboard for signal visualization; and 
(3) enabling scientific innovation via a command-line interface (CLI) that supports integration with machine learning and scalable workflow systems.

%%%%%%%%%%%%%%%%%%%%%%%%%%%%
\section{The CDMS Detectors}

Dark matter detectors are composed of three components: the target, the sensors, and the readout electronics. 
These components work in concert to detect rare interactions between dark matter and normal matter by capturing resulting signals such as heat, charge, or light.

%%%%%%%%%%%%%
{\bf Target: }
The target is the core material of the detector, like a crystal of germanium or silicon, where a dark matter particle is expected to interact, making it the first point of contact in the detection process. 
This paper focuses on Cryogenic Dark Matter Search (CDMS) detectors, which use silicon and germanium crystals as target materials. 
These materials are chosen because their relatively low nuclear masses are well-matched to potential dark matter candidates, maximizing interaction sensitivity. 
Since the mass of dark matter is unknown, different experiments use various target materials to explore a broad range of possibilities. CDMS is optimized for detecting dark matter particles with masses similar to nuclei and is particularly effective due to its moderate-sized target and well-characterized radioactive backgrounds.

\begin{figure}[!ht]
    \centering
    \includegraphics[width=1.0\linewidth]{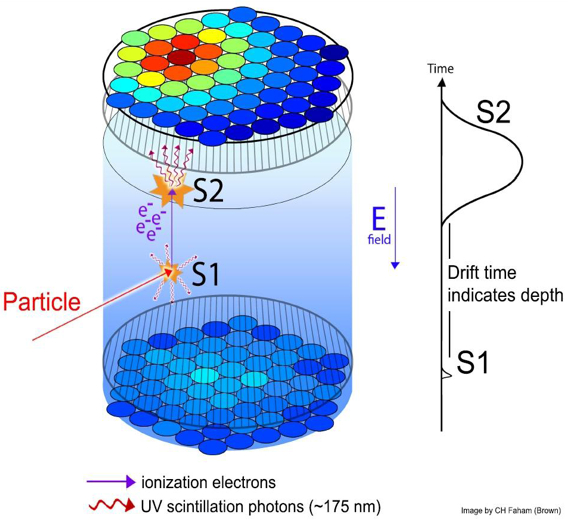}
    \caption{The Xenon experiment uses a wire grid to detect charge signals and an array of individual phototubes to detect light. Like CDMS, many dark matter experiments employ two distinct types of sensors and record their outputs as time-series data. Reprinted from~\cite{AKERIB2013111} with permission.}
    \label{fig:TPC}
\end{figure}

\begin{figure}[!ht]
    \centering
    \begin{subfigure}{0.45\textwidth}
        \centering
        \includegraphics[width=\linewidth]{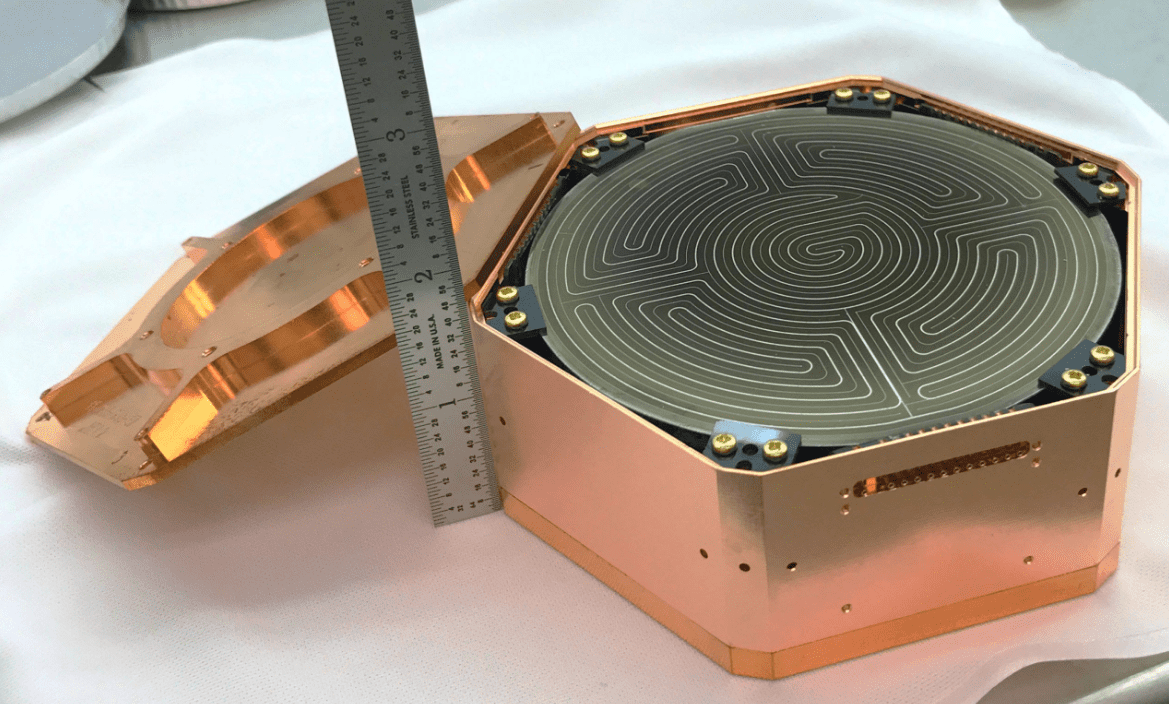}
        \caption{Photograph of a CDMS detector (from~\cite{physics-world2018}). The five darker regions are instrumented with phonon sensors, while the narrower regions in between contain charge sensors. Both the top and bottom surfaces of the germanium crystal are fully instrumented.}
        \label{fig:cdms-detector}
    \end{subfigure}
    \hfill
    \begin{subfigure}{0.5\textwidth}
        \centering
        \includegraphics[width=0.6\textwidth]{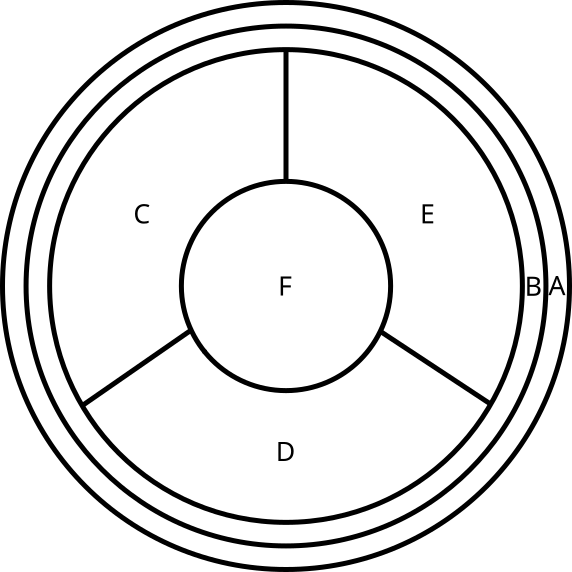}
        \caption{Schematic layout of the phonon sensors on the CDMS detector used at UMN. Each labeled region (A–F) represents an individual phonon channel.}
        \label{fig:channel-layout}
    \end{subfigure}
    \caption{CDMS detector layout: (a) A photograph showing the germanium target and surface-deposited sensors. The copper housing supports the crystal and routes readout cables. (b) A schematic of the phonon sensor arrangement used in the UMN dataset. Note that the layouts shown in (a) and (b) differ slightly.}
    \label{fig:cdms}
\end{figure}

%%%%%%%%%%%%%%%%%%%%%%%%%%
{\bf Sensor Technologies:} Sensors are attached to or embedded in the target material and detect physical signals, such as heat, charge, or light, produced by potential dark matter interactions. 
Detecting two distinct types of signals enables researchers to distinguish between electron recoils, typically caused by background radiation, and nuclear recoils, which may indicate dark matter interactions. 
The ratio of heat, charge, and light signals differ depending on the type of particle interaction, enabling effective background discrimination. 
Measuring only one signal would not provide enough information to make this distinction, while incorporating additional signals is often unnecessary or impractical due to material and design constraints.
Different experiments achieve this dual-signal detection in different ways. “Liquid noble” experiments (e.g., those using liquid xenon) rely on scintillation light and ionization charge (Figure~\ref{fig:TPC}). 
In contrast, CDMS detectors, targeted in this work, use germanium crystals, which do not scintillate and therefore cannot emit light.
Instead, CDMS detectors measure ionization charge and phonons (heat) generated during a particle interaction.
As shown in Figure~\ref{fig:cdms-detector}, the CDMS sensors are printed on both surfaces of the germanium crystal to collect these signals. 
Six phonon sensors (labeled A–F) are arranged in concentric and radial regions (Figure~\ref{fig:channel-layout}), interleaved with charge sensors. 
This configuration enables the simultaneous measurement of lattice vibrations (phonons or heat) and charge, providing key discrimination power between signal and background.

\begin{figure}[!ht]
    \centering
    \includegraphics[width=1\linewidth]{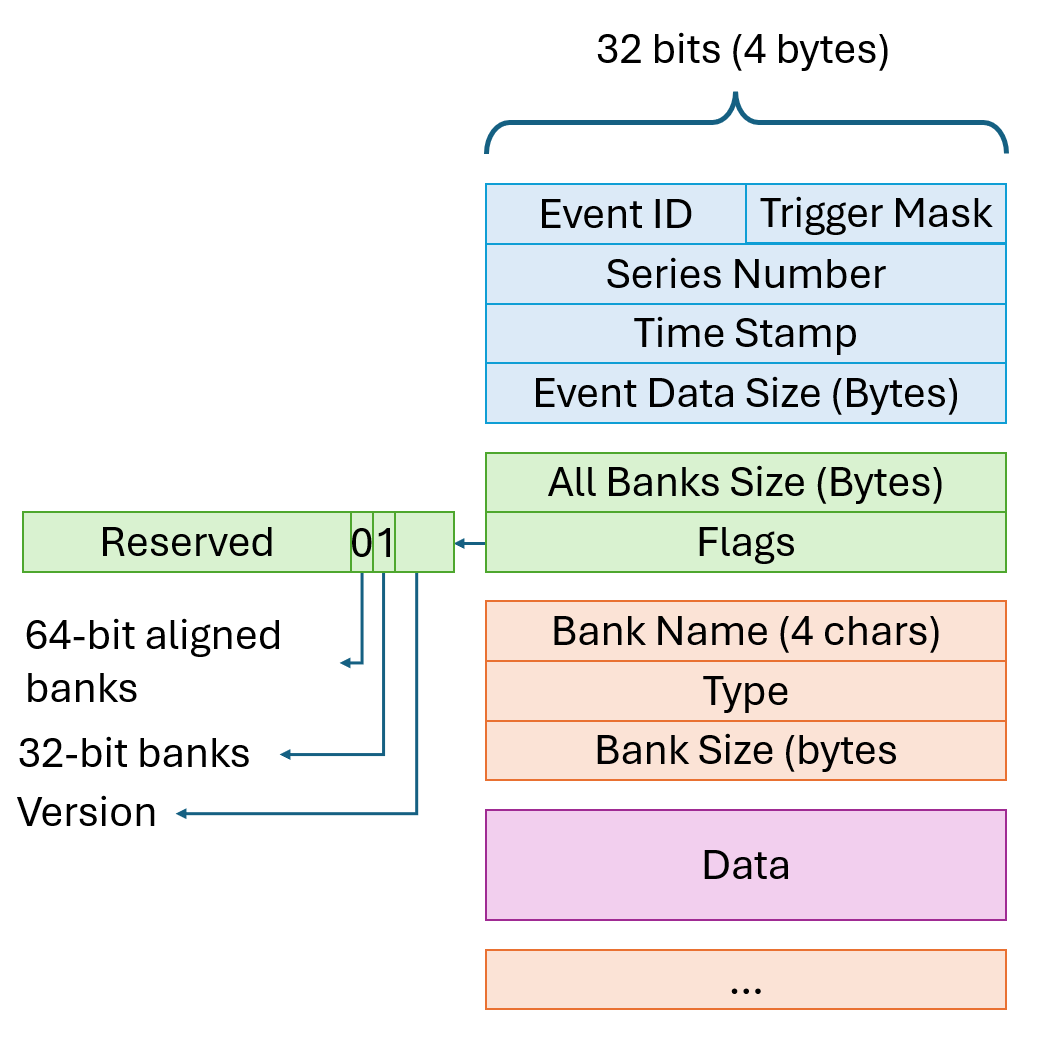}
    \caption{The MIDAS event format.  The data bank contains the CDMS data payload, which has its tightly packed format. Modified with permission from~\cite{midaswiki}}
    \label{fig:MEHBH}
\end{figure}

%%%%%%%%%%%%%%%%%%%%%%%%%%
{\bf Readout Electronics: } 
Sensor signals are analog and must be digitized to be stored and processed for analysis. 
Readout electronics digitize the signals generated by the sensors and pass them to data acquisition systems for recording and analysis.
CDMS uses a custom readout system connected to the MIDAS (Maximum Integration Data Acquisition System) framework, maintained by TRIUMF~\cite{triumf}. MIDAS organizes digitized signals into structured ``events" for each detector.
As illustrated in Figures~\ref{fig:MEHBH}, each MIDAS event contains multiple data banks and the ``Data" bank holds the raw CDMS payload. This payload includes time-series traces from phonon and charge channels. In the dataset used in this study, only the phonon traces are analyzed due to their superior signal-to-noise ratio.
% %%% 
% \begin{figure}[ht]
%     \centering
%     \includegraphics[width=1\linewidth]{MIDAS_CDMS_format.jpg}
%     \caption{The CDMS data payload.  This data is what goes in the pink "Data" box in the MIDAS event format.  This data format records digitized traces from phonon and charge channels for a set of detectors.  
%     %% The charge traces are not used in the UMN dataset and are not shown on the graphs.{\color{red} Need to request permission to use this and likely offer authorship to Belina and Ben!  }
%     }
%     \label{fig:MIDAS_CDMS_format}
% \end{figure}

%%%%%%%%%%%%%%%%%%%%%%%%%
\section{The R76 Dataset}

While most dark matter searches operate detectors deep underground to minimize background radiation, these environments are expensive and difficult to access.
In contrast, above-ground testing provides easier access and is commonly used for detector characterization and calibration. 
Between March and December 2022, a team at the University of Minnesota used a CDMS detector to explore a novel calibration technique in an above-ground setting~\cite{cushman2024strategiesmachinelearningapplied}. 
The resulting dataset, referred to as R76, has been made publicly available due to its ongoing scientific value.

\begin{figure*}[!ht]
    \centering
    \includegraphics[width=0.8\linewidth]{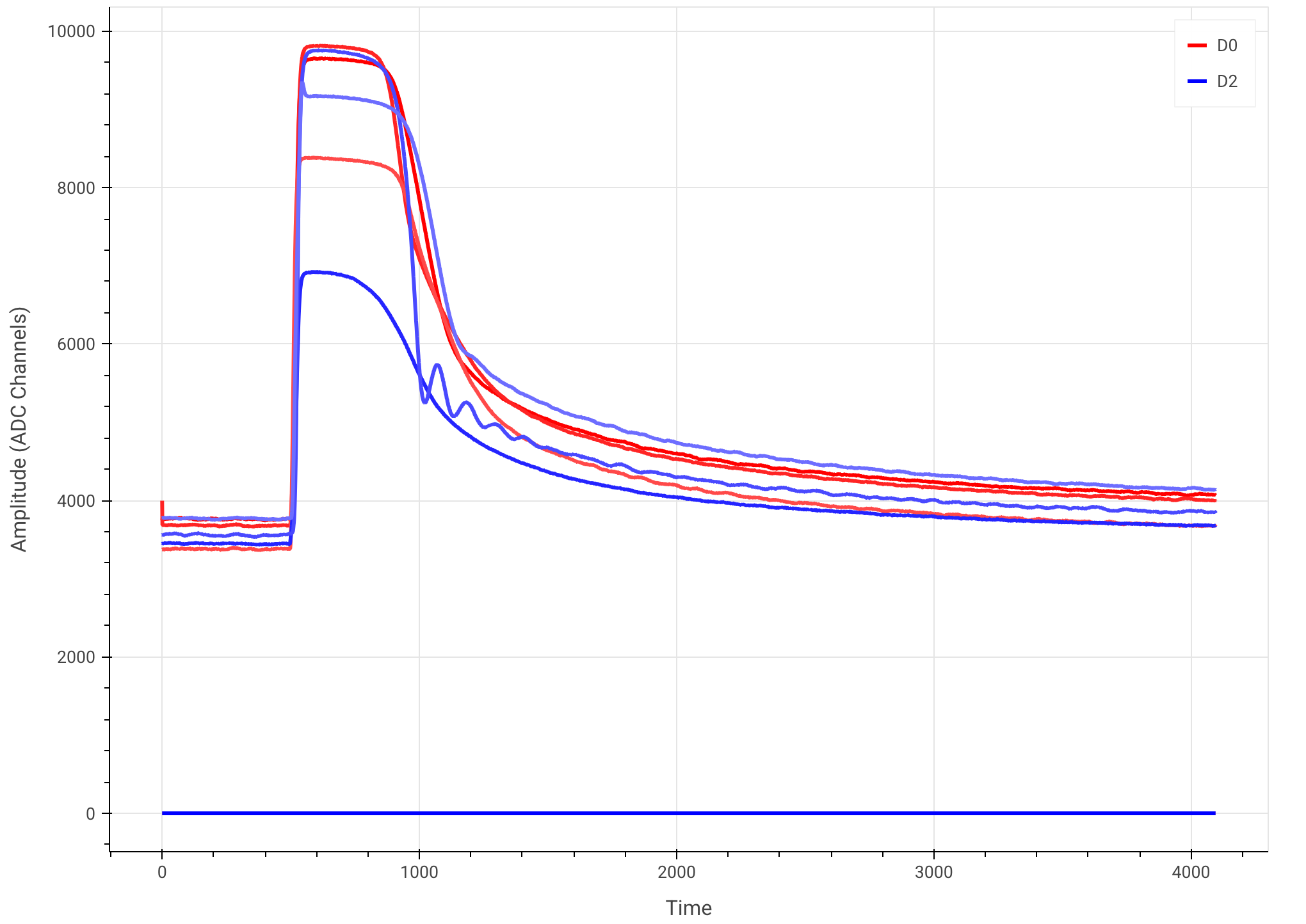}
    \caption{Example of a CDMS event showing the response of all channels on a single germanium crystal, viewed on the NSDF dashboard. The red and blue colors correspond to two digitizer readouts. The flat-topped pulses indicate signal saturation, characteristic of large-energy events useful for calibration rather than dark matter detection.}
    \label{fig:channel-response}
\end{figure*}

Unlike underground datasets intended for dark matter searches, R76 does not include data from the charge sensors. This is typical during calibration runs, where the focus is on phonon signals due to their superior signal-to-noise ratio. Figure~\ref{fig:channel-response} shows an example event from the dataset, where each trace represents a phonon channel on the detector. The flat-topped pulse shapes suggest signal saturation from a high-energy event, typical of calibration rather than a candidate dark matter interaction.

To enable detailed calibration, the experiment combined the CDMS detector with sodium iodide (NaI) scintillators. These NaI detectors are widely used in nuclear physics and were positioned to detect gamma-ray emissions coincident with nuclear recoils in the CDMS detector. 
This coincidence-based strategy enables the generation of a low-energy recoil spectrum with identifiable peaks, which is particularly valuable for calibrating the detector below 1 keV, a crucial challenge in dark matter experiments.

The team exposed the detector to three radioactive sources—$^{22}$Na, PuBe, and $^{241}$Am—across several shielding configurations:
\begin{itemize}
    \item $^{22}$Na was used with polyethylene and external lead shielding, although no shielding was placed between the source and the detector.
    \item PuBe, a strong thermal neutron source, was used with polyethylene shielding, full lead shielding around the detector, and additional borax shielding around the NaI array in later runs.
    \item $^{241}$Am was tested under all of the above configurations.
\end{itemize}
Each source was chosen for its ability to induce characteristic interactions: $^{22}$Na emits positrons and gammas; PuBe produces neutrons; and $^{241}$Am emits alpha particles and gammas.

The specific combinations of sources and shielding used in each configuration are summarized in Table~\ref{tab:configs}. The {\bf Basic} shielding refers to a configuration that includes polyethylene around the source and external lead shielding surrounding the detector setup, but not between the source and the detector. All other shielding configurations build on this baseline by adding additional materials, such as internal lead or borax. Specifically, {\bf Lead Bricks} indicates the addition of internal lead shielding, while {\bf Lead + Borax} refers to configurations that further include borax shielding, primarily to moderate neutrons around the NaI array.

\begin{table}[!h] 
    \centering
    \caption{Sources and shielding configurations used during the R76 calibration experiment.}
    \label{tab:configs}
    \begin{tabular}{|c|c|c|c|}
        \cline{2-4}
        \multicolumn{1}{c|}{} & $^{22}$Na & PuBe & $^{241}$Am \\ \hline
        Basic       & X &            & X \\ \hline
        Lead Bricks &            & X & X \\ \hline
        Lead + Borax &    & X & X \\ \hline
    \end{tabular}
\end{table}

%% Data value
As a result of these runs, the dataset contains a rich set of signals: many from background radiation and others from the de-excitation of nuclei that absorbed neutrons during source exposure. These signatures, particularly when coincident with CDMS and NaI detectors, provide a valuable resource for studying low-energy calibration and for testing analysis techniques in dark matter detection.

%%%%%%%%%%%%%%%%%%%%%%%%%%%%%%%%%
\section{Unlocking R76 with NSDF}

Despite the rich utility of the R76 dataset, we encountered three challenges that limited its reach:
(1) lack of access through open-source tools; 
(2) a steep, time-consuming learning curve; and
(3) limited support for data-driven innovation. 
These challenges are not intrinsic to the dataset itself but are symptoms of the siloed nature of data tools within the dark matter community.
To address these challenges, we take advantage of the National Science Data Fabric (NSDF)~\cite{nsdf_services}, a cyberinfrastructure ecosystem designed to streamline scientific data management. 
The need for a robust, accessible data fabric has grown as scientific datasets expand in size and complexity. Traditionally, researchers have had to navigate a patchwork of computing providers, limited funding for software development, and domain-specific technical requirements. 
This fragmented landscape can hinder collaboration and slow scientific progress, particularly for large-scale datasets requiring significant computational resources for storage, processing, and visualization.

NSDF bridges these gaps by providing services~\cite{nsdf_services} that integrate networking~\cite{nsdf_network}, storage~\cite{nsdf_cloud, nsdf_fair_objects}, workforce development ~\cite{nsdf_tutorial}, and computing resources on academic and commercial cloud infrastructures ~\cite{nsdf_cloud, nsdf_fuse}. 
By coordinating data movement between geographically distributed teams and enabling scalable and interoperable workflows, NSDF lowers barriers to cloud-based cyberinfrastructure. 
Additionally, NSDF facilitates the adoption of FAIR (Findable, Accessible, Interoperable, Reusable) data principles, ensuring that scientific datasets remain discoverable and usable across disciplines.

\subsection{Increasing Data Accessibility with Open-Source Tools: The MIDAS-to-IDX Transition }

Dark matter experiments often rely on custom data formats and monolithic analysis software developed within individual collaborations. 
For instance, the original CDMS dataset is recorded in the MIDAS format, a collaboration-specific structure that presents challenges for broader accessibility due to its tightly coupled encoding.
As a result, tools across the community are fragmented, poorly documented, and difficult to use outside their original context. 
This fragmentation creates significant barriers to data reuse and innovation, particularly for new researchers or interdisciplinary collaborators.

The NSDF platform addresses this challenge by providing a data transformation service that converts specific data formats into widely supported structures. 
For the CDMS dataset, we use NSDF services to convert data from the MIDAS format into the IDX format, a layout widely adopted in other scientific domains. 
This transformation improves accessibility and supports real-time exploration through visual dashboards and scalable analytics frameworks~\cite{10767643,11044452,kumar2010towards}.

The conversion process begins with parsing the MIDAS files using the CDMS-specific IOLibrary, which extracts digitized traces from the phonon channels. 
These traces are stored in an intermediate NumPy Zipped (NPZ) format, preserving event structure, detector metadata, and channel assignments. 
The NPZ data is then passed to a conversion tool built with the OpenVisusPy library, which generates three essential output files:
\begin{itemize}
    \item an \textbf{IDX file} containing the multi-resolution, hierarchically indexed channel data;
    \item a \textbf{TXT file} mapping channel data to detectors, events, and channels; and
    \item a \textbf{metadata file} capturing key experimental parameters such as trigger type, readout configuration, and timestamps.
\end{itemize}
The IDX layout preserves the temporal and spatial relationships of the original MIDAS data and adds efficient indexing for slicing and subsetting. 
It also supports multi-resolution encoding, allowing users to stream low-resolution previews and progressively refine to full resolution. 
This cache-oblivious approach enables scalable interaction with large datasets without overloading memory.
Although HDF5~\cite{hdf5} and TIFF~\cite{tiff} formats are effective for archival storage and detailed analysis, the IDX format has demonstrated strong performance for web-based progressive visualization~\cite{10767643}, making it well suited for exploratory dark matter research.

\begin{figure}[ht]
    \centering
    \includegraphics[width=0.8\linewidth]{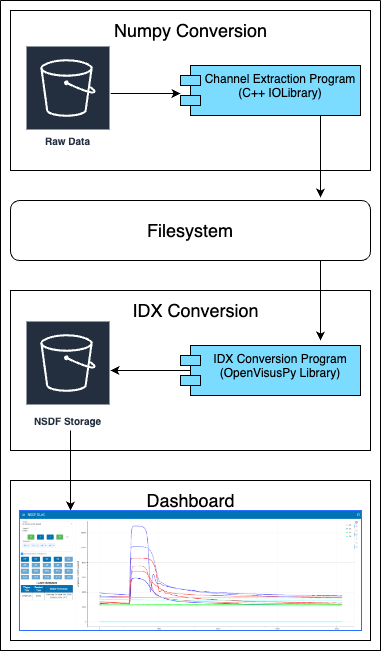}
    \caption{Workflow for converting MIDAS data into the IDX format using NSDF services. The process includes signal extraction via IOLibrary, intermediate NPZ storage, hierarchical conversion via OpenVisusPy, and packaging for persistent cloud-based access.}
    \label{fig:idx_conversion}
\end{figure}

To ensure persistence and accessibility, the IDX, TXT, and metadata files are uploaded to NSDF storage services, which support data collections ranging from terabytes to petabytes.
%%%
Figure~\ref{fig:idx_conversion} shows the full workflow for converting the CDMS data from MIDAS to IDX format, including intermediate steps and tools used.

%%%%%%%%%%%%%%%%%%%%%%%%%%%%%%%%%%%%%%%%%%%%%%%%%%%%%%%%%%%%%%%%%%%% 
\subsection{Shortening the Learning Curve: A Web-Based Dashboard for Interactive Signal Visualization }

The steep learning curve associated with the CDMS collaboration’s legacy software presents a significant barrier to interdisciplinary research and workforce training. 
New users, especially those outside of the collaboration, must invest years to become proficient in the software stack. 
A few graduate students and postdoctoral researchers do achieve expertise and become indispensable, but their knowledge is difficult to transfer, and the collaboration often lacks the resources to improve the software’s usability. 
As a result, potential collaborators are frequently discouraged by the multi-year ramp-up required to access and analyze the data.

\begin{figure*}[!ht]
    \centering
    \includegraphics[width=\linewidth]{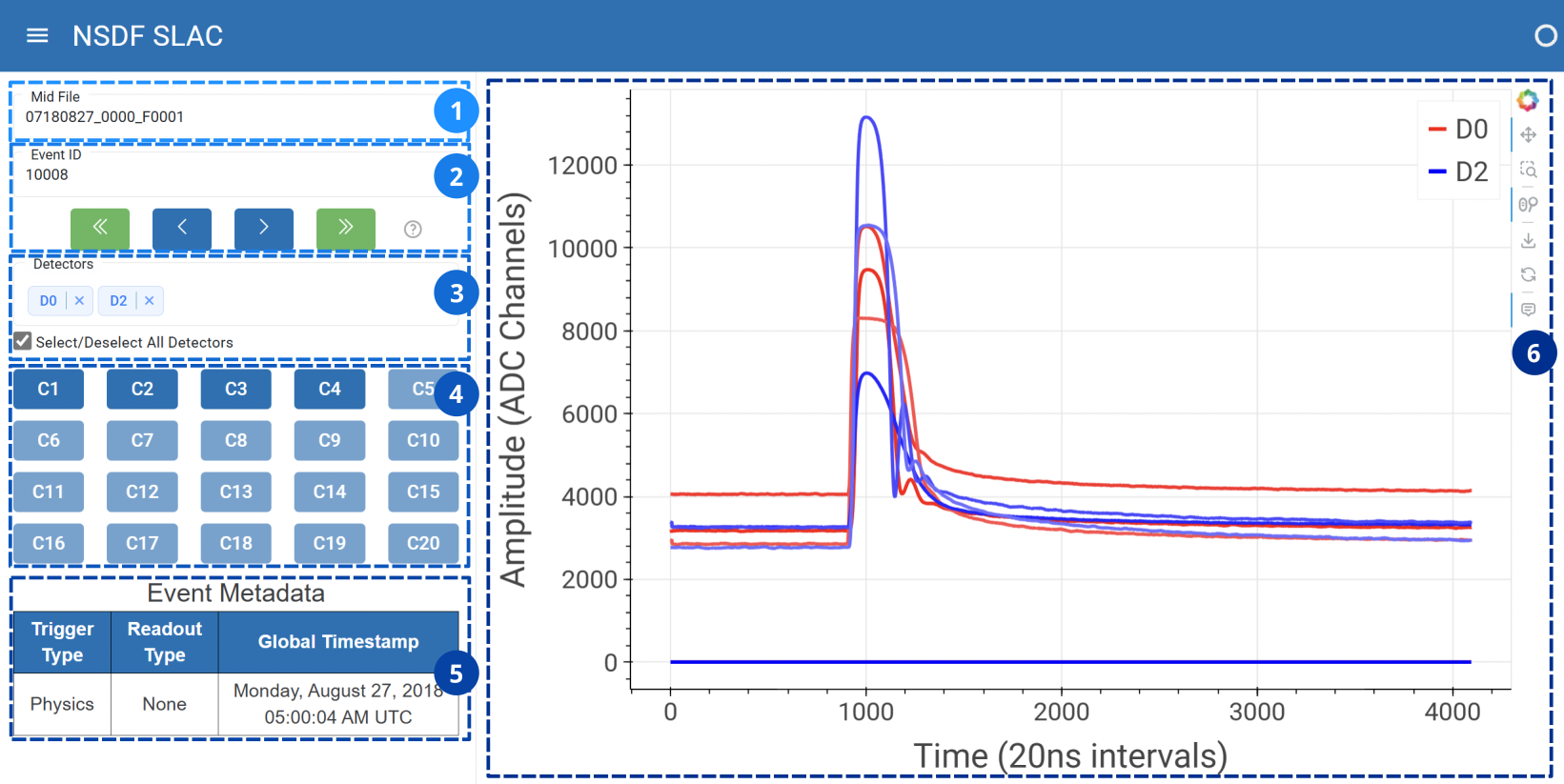}
    \caption{The NSDF dashboard displays CDMS data in a web browser. Control (1) allows file selection and control (2) allows event selection. Control (3) filters data by electronics readout boards, and control (4) selects individual channels. Table (5) shows event metadata, and plot (6) provides an interactive view of pulse shapes.}
    \label{fig:dashboard}
\end{figure*}

To address this barrier, NSDF developed a web-based dashboard that dramatically reduces the learning curve for accessing and exploring CDMS data. The platform provides a highly interactive and responsive environment for visualizing the converted IDX signals, as shown in Figure~\ref{fig:channel-response} and Figure~\ref{fig:dashboard}. By abstracting the complexities of legacy data formats, the dashboard enables immediate involvement with the dataset for students, researchers, and interdisciplinary teams.

Built with modern web technologies and Python libraries such as Panel~\cite{panel_library}, alongside JavaScript frameworks, the dashboard supports intuitive exploration of event-level data. 
Researchers can inspect low-level signal traces, identify interesting patterns, and access detailed metadata—all within a user-friendly interface. 
This accessibility has been particularly valuable for onboarding new students and fostering cross-disciplinary collaborations. 
The dashboard is also fully integrated into the broader NSDF ecosystem, allowing seamless access to NSDF-managed storage and compute services.
%%%
The interface provides the following core features, as illustrated in Figure~\ref{fig:dashboard}:
\begin{enumerate}
    \item \textit{MID file selection:} Choose specific .mid files from storage for analysis.
    \item \textit{Event navigation:} Enter event IDs to retrieve and examine individual events.
    \item \textit{Detector selection:} Filter the data by selecting specific detectors.
    \item \textit{Channel visualization:} Toggle individual channels on or off using a grid of buttons.
    \item \textit{Event metadata display:} View a table with detailed metadata for each selected event.
    \item \textit{Interactive plotting:} Dynamically explore signal traces using an automatically updating central plot area.
\end{enumerate}

%%%%%%%%%%%%%%%%%%%%%%%%%%%%%%%%%%%%%%%%%%%%%%%%%%%%%%%%%%%%%%%%%%%%%
\subsection{Enabling Science Innovation: Interoperable Tools for Dark Matter Analysis}

Although the R76 dataset has been publicly available for over a year, only members of the CDMS collaboration have been able to use it effectively. This limitation stems from the need to master a private, complex, and monolithic software stack. As a result, scientific innovation has been significantly hindered. For example, despite the surge in interest around using machine learning to improve position and energy estimates, integrating modern ML libraries with CDMS data remains prohibitively difficult due to software and access barriers.

The NSDF platform addresses this challenge by enabling the integration of new tools into its ecosystem and supporting their deployment across scalable infrastructure. 
Leveraging this platform, we developed a suite of interoperable tools that allow researchers to access and analyze IDX-formatted CDMS data without relying on the original monolithic software stack. 
These tools support both (1) high-throughput file-level data retrieval and (2) intuitive data selection at the event and channel level, returning standard NumPy arrays that integrate seamlessly with Python libraries such as TensorFlow.

NSDF delivers these tools through a robust Command-Line Interface (CLI) built using the Typer framework~\cite{typer}. 
Designed for flexibility and ease of use, the CLI allows users with varying technical expertise to incorporate the tools into automated workflow systems such as Pegasus and Snakemake with minimal overhead.

To support scalable, reproducible analysis pipelines, the NSDF tools are designed to meet several key requirements:

\paragraph{Metadata-based file access. } 
The tools abstract the intricacies of remote data storage and allow researchers to download specific subsets of data based on various metadata attributes. These files are typically downloaded to a high-performance computing environment where the workflow software is run.

\paragraph{High-throughput performance. } 
Even the most extensive datasets must be readily available for analysis. 
The tools are optimized for speed, enabling rapid retrieval of large volumes of data without creating bottlenecks. This feature is critical for workflows that train ML models or perform iterative scans across event data.

\paragraph{Customizable event filtering. }  
CDMS data consists of physics “events,” and each event includes data for each channel on each detector. It is common for analyzers to need access to specific detectors or channels. 
The tools allow users to specify exactly which events, detectors, and channels they want.

\paragraph{Python ecosystem integration. } 
To support modern scientific analysis, the tools are designed to deliver data as standard NumPy arrays. This format ensures seamless compatibility with Python-based libraries and frameworks. For example, the R76 dataset is currently being used to train a Generative Adversarial Network (GAN) that generates synthetic signals to validate dark matter detection methods. By removing the need for specialized CDMS software, these tools empower researchers to develop, share, and reproduce algorithms more easily across the community.

A representative use case of this NSDF-enabled toolchain involves an under-development ML workflow. 
An interdisciplinary team first uses the NSDF dashboard to visually inspect and select relevant MID files. The CLI then retrieves those files to an HPC system where the GAN is trained—using nearly a terabyte of memory, far exceeding the capacity of typical end-user machines. 
The entire workflow is orchestrated using Pegasus~\cite{deelman2015pegasus}, ensuring reproducibility across compute environments and simplifying reusability for outside collaborators.

%%%%%%%%%%%%%%%%
\section{Conclusion and Outlook}

Dark matter detectors like CDMS are custom-designed for low thresholds and minimal background radioactivity, offering unique insight into one of the most compelling open questions in physics. Yet despite their scientific value, these datasets remain underutilized due to several barriers: custom data formats not compatible with open-source tools, steep learning curves for collaboration-specific software, and limited opportunities for innovation beyond the original research teams.
This paper presents how the National Science Data Fabric (NSDF) overcomes these limitations using the publicly available R76 calibration dataset from the University of Minnesota. 
NSDF enables three key capabilities: (1) transformation of collaboration-specific MIDAS data into IDX format for scalable visualization and analysis; (2) a web-based dashboard for interactive data exploration and onboarding; and (3) a command-line interface (CLI) that integrates seamlessly with Python libraries and workflow systems such as Pegasus.
Together, these services lower the technical barriers to entry and open new avenues for interdisciplinary collaboration, allowing researchers from physics, computer science, and data science to work directly with complex dark matter data.

Looking ahead, NSDF will continue to expand support for new data formats, visualization frameworks, and scalable analysis pipelines, further democratizing access to dark matter science through automation, reproducibility, and interoperability.
\textit{We invite the broader research community to explore the R76 dataset using NSDF tools, contribute improvements, and develop new scientific insights.}

\section{ACKNOWLEDGMENT}

This research is supported by the National Science Foundation (NSF) awards \#2138811, \#2103845, \#2334945, \#2138296, \#2331152, \#2104003, \#2127548,  \#2330582 and \#2017760.   Additionally, this research is supported in part by the Advanced Research Projects Agency for Health (ARPA-H) grant no. D24AC00338-00, the Intel oneAPI Centers of Excellence at University of Utah, the NASA AMES cooperative agreements 80NSSC23M0013 and NASA JPL Subcontract No. 1685389 and by the Department of Energy (DOE) grant DE-SC0023898. The work presented in this paper was partly obtained using resources from ACCESS TG-CIS210128 and ACCESS PHY210008. We thank the Dataverse and ScientistCloud by ViSOAR LLC.

\begin{paragraph}{Amy Roberts}{\,} is an Assistant Professor in the Physics department at the University of Colorado Denver.  She earned her Ph.D. in experimental nuclear physics at the University of Notre Dame in 2013 after graduating with her B.Sc. in both Physics and Mathematics from the State University of New York - Stony Brook.  Her research interests focus on making dark matter discovery possible with the next generation of dark matter detectors by building community trust in analysis.
\end{paragraph}

\begin{paragraph}{Jack Marquez}{\,} is a Research Assistant Professor at the University of Tennessee, Knoxville, as a member of the Global Computing Laboratory directed by Dr. Michela Taufer. He earned his Ph.D. in Engineering in 2022 from the University Autonoma de Occidente (Cali, Colombia) after graduating with his BSc in Informatics Engineering from the same university in 2013. His research interests include HPC, Cloud Computing, heterogeneous storage systems, and performance optimization.
\end{paragraph}

\begin{paragraph}{Kin Hong NG}{\,} is a Research Scientist at the University of Tennessee, Knoxville, as a member of the Global Computing Laboratory directed by Dr. Michela Taufer. He earned his BSc in Computer Science from the University of South Florida (USF) in 2021. His research focuses on the processing, containerization, and visualization of data for scientific workflows.
\end{paragraph}

\begin{paragraph}{Kitty Mickelson}{\,} completed their Master's in Integrated Sciences degree in the areas of Physics and Mathematics at the University of Colorado Denver in 2025. Their research focuses on detector calibration using nuclear physics concepts, probabilistic modeling, and statistical analysis. 
\end{paragraph}

\begin{paragraph}{Aashish Panta}{\,} is a PhD student at the University of Utah and a Graduate Research Assistant at the Scientific Imaging and Computing Institute. His research interests include large-scale data management, visualization techniques, and leveraging machine learning algorithms on large datasets.
\end{paragraph}

\begin{paragraph}{Giorgio Scorzelli}{\,} is the director of software development at the Center for Extreme Data Management Analysis and Visualization, University of Utah, Salt Lake City, UT,
84112, USA, and software development director for the National Data Science Fabric. 
\end{paragraph}

\begin{paragraph}{Amy Gooch}{\,}Chief Operating Officer at ViSOAR LLC and a member of the University of Utah’s Scientific Computing and Imaging Institute, she spearheads advanced storage-as-a-service solutions for scientific research, focusing on large-scale data management and visualization to facilitate efficient analysis and collaboration across various scientific domains.
\end{paragraph}

\begin{paragraph}{Priscilla B. Cushman}{\,} is a Professor of Physics at the University of Minnesota, Minneapolis, and the spokesperson of the CDMS experiment.  She earned her Ph.D. at Rutgers University.  Her research focuses on Particle Physics, direct detection of dark matter, and cryogenic phonon detectors.
\end{paragraph}

\begin{paragraph}{Matt Fritts}{\,} is a Senior Staff Scientist at Washington University in Saint Louis.  He got his PhD at the University of Minnesota, Minneapolis.  His research interests are
particle physics, cryogenic detectors, dark matter direct detection, and gamma-ray astronomy.
\end{paragraph}

\begin{paragraph}{Himangshu Neog}{\,} is a postdoctoral scholar at the University of Minnesota, Minneapolis, and is a member of the CDMS experiment.  He earned his Ph.D. at Texas A\&M University.  His research focuses on direct detection of dark matter, dilution refrigerators, and cryogenic phonon detectors.
\end{paragraph}

\begin{paragraph}{Valerio Pascucci}{\,} is the Inaugural John R. Parks Endowed Chair; the founding director of the Center for Extreme Data Management Analysis and Visualization, a faculty of the Scientific Computing and Imaging Institute; and a professor in the School of Computing at the University of Utah, Salt Lake City, UT, 84112, USA. He is the principal investigator for the National Data Science Fabric.
\end{paragraph}

\begin{paragraph}{Michela Taufer}{\,} holds the Jack Dongarra Professorship in High-Performance Computing within the Department of Electrical Engineering and Computer Science at the University of Tennessee, Knoxville. Dr. Taufer received her Ph.D. in computer science from the Swiss Federal Institute of Technology (EHT) in 2002. 
Her interdisciplinary research is at the intersection of computational sciences, high-performance computing, and data analytics. 
\end{paragraph}

\balance
\bibliographystyle{IEEEtran}
\bibliography{references}

\end{document}